\documentclass{iopart}

\usepackage{graphicx}

\begin{document}
\title[Homodyne detection and optical parametric amplification]{Homodyne detection and optical 
parametric amplification: a classical approach applied to proposed ``loophole-free'' 
Bell tests}

\author{Caroline H Thompson\footnote{Email: ch.thompson1@virgin.net}}

\address{11 Parc Ffos, Ffos-y-Ffin, Aberaeron, SA46 0HS, U.K.}

\date{\today}

\begin{abstract}

Recent proposed ``loophole-free'' Bell tests are discussed in the light of 
classical models for the relevant features of optical parametric amplification
and homodyne detection.  The Bell tests themselves are uncontroversial: 
there are no obvious loopholes that might cause bias and hence, if the world 
does, after all, obey local realism, no violation of a Bell inequality will 
be observed. Interest centres around the question of whether or not the 
proposed criterion for ``non-classical'' light is valid. If it is not, then 
the experiments will fail in their initial concept, since both quantum 
theorists and local realists will agree that we are seeing a purely 
classical effect. The Bell test, though, is not the only criterion by which 
the quantum-mechanical and local realist models can be judged. It is 
suggested that the quantum-mechanical models given in the proposals will also 
fail in their detailed predictions.  If the experiments are extended by 
including a range of parameter values and by analysing, in addition to the 
proposed digitised voltage differences, the raw voltages, the models can be 
compared in their overall performance and plausibility.
\end{abstract}

%\pacs{03.65.Ud, 03.67.-a, 03.67.Mn, 42.25.-p, 42.50.Dv, 42.50.Xa}
				% PACS, the Physics and Astronomy Classification Scheme.
 %List as amended by PRA editor
\vspace{2pc}


\submitto{\JOB}

\maketitle

\section{Introduction}

No test of Bell's inequalities \cite{Bell:1964, Bell:1971} to date has been free 
of ``loopholes''. This means that, despite the high levels of statistical 
significance frequently achieved, violations of the inequalities could be the 
effects of experimental bias of one kind or another, not evidence for the presence of 
quantum entanglement. Recent proposed experiments by Garc\'{\i}a-Patr\'{o}n 
\etal~\cite{Garc:2004} and Nha and Carmichael \cite{Nha:2004} show 
promise of being genuinely free from such problems. If the world in fact 
obeys local realism, they should \textit{not}, therefore, infringe any Bell inequality.

\begin{figure}[htbp]
\centerline{\includegraphics[width=2.6in,height=2.6in]{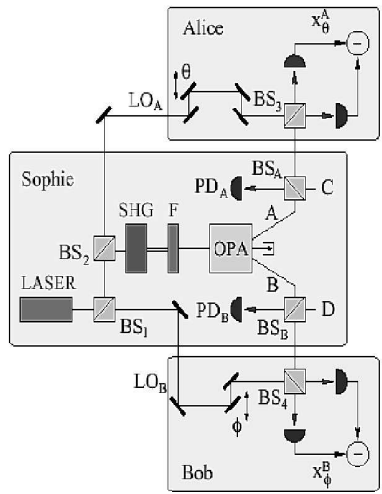}}
\caption{Proposed experimental set-up. 
In the current text, phase shifts $\theta $ and $\phi $ are renamed $\theta 
_{A}$ and $\theta _{B}$. For further explanation, see main text. 
(\textit{Reprinted with permission from Garc\'{\i}a-Patr\'{o}n et al., 
Phys.~Rev.~Lett.~}\textbf{\textit{93}}\textit{, 130409 (2004). 
Copyright (2004) by the American Physical Society.
})}
\label{fig1}
\end{figure}

The current article discusses a classical model that should be able, once all 
relevant details are known, to explain the results, accounting not only for the failure to 
infringe the selected Bell test but also for other failures in the detailed predictions. 
It depends on the classical theory for homodyne detection (re-derived here from first
principles) and the known behaviour of (degenerate-mode) optical parametric 
amplifiers (OPA) \cite{Walls:1994}. 

As far as the ``loophole-free'' status of the proposed experiments is 
concerned, there would appear to be no problem. A difficulty that seems 
likely to arise, though, is that theorists may not agree that the test beams 
used were in fact ``non-classical'', so the failure to infringe a Bell 
inequality will not in itself be interpreted as showing a failure of quantum 
mechanics\footnote{
The predicted violation is in any case small, so failure may be put 
down to other ``experimental imperfections''.}.  
The criterion to be used to establish the non-classical nature of the light is 
the observation of negative values of the Wigner density, and there is reason 
to think that, even if the standard method of estimation seems to show that 
these are achieved, the method may be in error. Wigner density is, 
in any event, irrelevant to our model.  Far from being, as suggested by 
Garc\'{\i}a-Patr\'{o}n and others, the ``hidden variable'' needed, it plays no 
part whatsoever. 

Regardless of the outcome of the Bell tests, and whether or not the light is 
declared to be non-classical, there are features of the experiments that can 
usefully be exploited to compare the strengths of quantum mechanics versus 
(updated) classical theory as viable models. The two theories approach the 
situation from very different angles. Classical theory traces the causal 
relationships between phenomena, starting with the simplest assumption and 
building in random factors later where necessary. Quantum mechanics starts 
with models of complete ensembles, all random factors included. This, it is 
argued, is inappropriate, since two features of the proposed experiments 
demand that we consider the behaviour of individual events, not whole 
ensembles: the process of homodyne detection itself, and the Bell test.

\section{The proposed experiments}

The experimental set-up proposed by Garc\'{\i}a-Patr\'{o}n \textit{et al.} is shown 
in Fig.~1, that of Nha and Carmichael being similar. In the words of the 
Garc\'{\i}a-Patr\'{o}n \textit{et al.}~proposal:

\begin{quotation}
\noindent
The source (controlled by Sophie) is based on a master laser beam, which 
is converted into second harmonic in a nonlinear crystal (SHG). After 
spectral filtering (F), the second harmonic beam pumps an optical parametric 
amplifier (OPA) which generates two-mode squeezed vacuum in modes A and B. 
Single photons are conditionally subtracted from modes A and B with the use 
of the beam splitters BS$_{A}$ and BS$_{B}$ and single-photon detectors 
PD$_{A}$ and PD$_{B}$. Alice (Bob) measures a quadrature of mode A (B) using 
a balanced homodyne detector that consists of a balanced beam splitter 
BS$_{3}$ (BS$_{4 })$ and a pair of highly-efficient photodiodes. The local 
oscillators LO$_{A}$ and LO$_{B}$ are extracted from the laser beam by means 
of two additional beam splitters BS$_{1}$ and BS$_{2}$.
\end{quotation}

The classical description, working from the same figure, is just a little 
different. Quantum theoretical terms such as ``squeezed vacuum'' and 
``quadrature''\footnote{The usual model for the electric field as the sum of two
orthogonal quadratures is appropriate where there is no base-line for the phase 
but not, as here, where all phases concerned are defined and measured relative to
a definite base, namely that of the master laser.  As will be seen, it it phase differences
of $180^{\circ}$, not $90^{\circ}$, that feature in the classical model.} 
 are not used since they are not appropriate to the model and would cause confusion. 
The description might run as follows:

The master laser beam (which is, incidentally, pulsed) is frequency-doubled 
in the crystal SHG. After filtering to remove the original frequency, the 
beam is used to pump the OPA, a resonant cavity containing a nonlinear crystal
cut so as to produce degenerate parametric down-conversion of the input.  The 
output comprises pairs of classical wave pulses at half the input frequency, 
i.e.~at the original laser frequency. 
The selection of pairs for analysis is done by splitting each output at an 
unbalanced beamsplitter (BS$_{A}$ or BS$_{B})$, the smaller parts going to 
sensitive detectors PD$_{A}$ or PD$_{B}$. Only if there are detections at 
both PD$_{A}$ and PD$_{B}$ is the corresponding homodyne detection included 
in the analysis. The larger parts proceed to balanced homodyne detectors, 
i.e.~ones in which the intensities of local oscillator and test inputs are 
approximately equal. The source of the local oscillators LO$_{A}$ and 
LO$_{B}$ is the same laser that stimulated, after frequency doubling, the 
production of the test beams. 

\section{Homodyne detection}
In (balanced) homodyne detection, the test beam is mixed at a beamsplitter 
with a local oscillator beam of the same frequency and the two outputs sent 
to photodetectors that produce voltage readings for every input pulse. In 
the proposed ``loophole-free'' Bell tests the difference between the two 
voltages will be converted into a digital signal by counting all positive 
values as +1, all negative as --1.

\begin{figure}[htbp]
\centerline{\includegraphics[width=1.5in,height=1.5in]{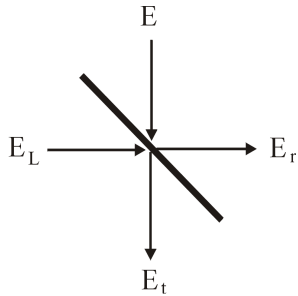}}
\caption{Inputs and outputs at the beamsplitter in a homodyne detector.
E$_{L }$ is the local oscillator beam, E the test beam, E$_{t }$ and E$_{r}$  
the transmitted and reflected beams respectively.}
\label{fig2}
\end{figure}

Assuming the inputs are all classical waves of the same frequency and there 
are no losses, it can be shown (see below) that the difference between the 
intensities of the two output beams is proportional to the product of the 
two input intensities multiplied by $\sin \theta$, where $\theta$ is the phase 
difference between the test beam and local oscillator. If voltages are 
proportional to intensities then it follows that the voltage difference will 
be proportional to $\sin \theta $. When digitised, this transforms to a step function, 
taking the value $-1$ for $-\pi < \theta < 0$ and $+1$ for $0 < \theta < \pi $.  
(The function is not well defined for integral multiples of $\pi$.)

\subsection{Classical derivation of the predicted voltage difference}
Assume the test and local oscillator signals have the same frequency, 
\textit{$\omega $}, the time-dependent part of the test signal being modelled by $e^{i\phi }$, 
where (ignoring a constant phase offset\footnote{The phase offset depends on the difference in optical path lengths, 
which will not in practice be exactly constant due to thermal oscillations. If a 
complete model is ever constructed, this should therefore be a parameter.})
\textit{$\phi =\omega $t} is the phase angle, 
and the local oscillator phase and test beam phases differ by \textit{$\theta $}. [Note that 
although complex notation is used here, only the real part has meaning: this 
is an ordinary wave equation, not a quantum-mechanical ``wave function''. To 
allay any doubts on this score, the derivation is partially repeated with no 
complex notation in the Appendix.]

\begin{figure}[htbp]
\centerline{\includegraphics[width=1.8in,height=1.8in]{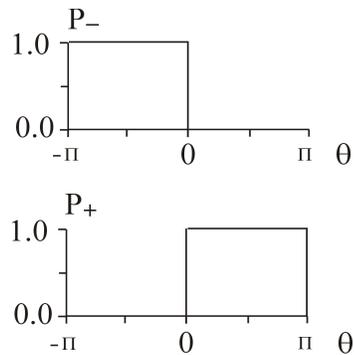}}
\caption{Probabilities of `+' and `--' outcomes versus phase difference, 
using digitised difference voltages from a perfect, noise-free, balanced 
homodyne detector.}
\label{fig3}
\end{figure}

Let the electric fields of the test signal, local oscillator and reflected and transmitted signals 
from the beamsplitter have amplitudes $E$, $E_{L}$, $E_{r}$ and $E_{t}$ respectively, as 
shown in Fig.~\ref{fig2}. Then, after allowance for phase delays of $\pi/2$ at each 
reflection and assuming no losses, we have
\begin{equation}
\label{eq1}
E_r = \frac{1}{\sqrt 2} (Ee^{i(\phi +\pi /2)}+E_L e^{i(\phi +\theta )})
\end{equation}
and
\begin{equation}
\label{eq2}
E_t = \frac{1}{\sqrt 2} (Ee^{i\phi }+E_L e^{i(\phi +\theta +\pi /2)}).
\end{equation}
The intensity of the reflected beam is therefore
\begin{eqnarray}
\label{eq3}
E_r E_r^\ast & = & \frac{1}{2} (Ee^{i(\phi +\pi /2)}+E_L e^{i(\phi +\theta )})(Ee^{-i(\phi 
+\pi /2)}
\nonumber \\
             & + & E_L e^{-i(\phi +\theta )})
\nonumber \\
             & = & \frac{1}{2}(E^2+E_L ^2+EE_L e^{i(\pi /2-\theta )}+EE_L e^{-i(\pi /2-\theta )})
\nonumber \\
             & = & \frac{1}{2} (E^2+E_L ^2+2EE_L \cos (\pi /2-\theta )
\nonumber \\
             & = & \frac{1}{2} (E^2+E_L ^2+2EE_L \sin \theta ).
\end{eqnarray}
Similarly, it can be shown that the intensity of the transmitted beam is
\begin{equation}
\label{eq4}
E_t E_t^\ast = \frac{1}{2} (E^2+E_L ^2-2EE_L \sin \theta ).
\end{equation}
If the voltages registered by the photodetectors are proportional to the 
intensities, it follows that the difference in voltage is proportional to 
$2EE_L \sin \theta$. When digitised, this translates to the 
step function mentioned above. The probabilities for the two possible 
outcomes are, as shown in Fig.~\ref{fig3},

\begin{equation}
\label{eq5}
p_{-} = \left\{
\begin{array}{ll}
     1 & \mbox{for $-\pi < \theta < 0$} \\
     0 & \mbox{for $0 < \theta < \pi$}
\end{array}
\right.
\end{equation}
and
\begin{equation}
\label{eq6}
p_{+} = \left\{
\begin{array}{ll}
     0 & \mbox{for $-\pi < \theta < 0$} \\
     1 & \mbox{for $0 < \theta < \pi$}
\end{array}
\right.
\end{equation}

Note that the probabilities are undefined for integral multiples of $\pi$.  
In practice it would be reasonable to assume that, due to the presence of noise,
 all the values were 0.5, but for the present purposes the integral values will
simply be ignored.

\section{Application to the proposed Bell tests}
If the frequencies and phases of both test beams and both local oscillators 
were all identical apart from the applied phase shifts, the experiments would 
be expected to produce step function relationships between counts and 
applied shifts both for the individual (singles) counts and for the 
coincidences. 

\begin{figure}[htbp]
\centerline{\includegraphics[width=2.7in,height=1.8in]{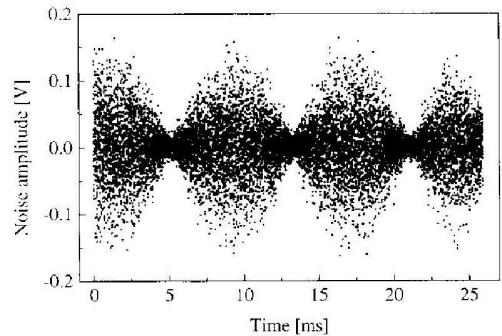}}
\caption{A typical scatter of ``noise amplitude'' (related to voltage difference)
 as phase $\theta$ is scanned over time.  Minimum scatter occurs for  $\theta$ taking 
integral multiples of $\pi$. (\textit{Reprinted from G. Breitenbach \etal,  
{\it J.~Opt.~Soc.~Am. B}~\textbf{12} 2304 (1995).})}
\label{fig4}
\end{figure}

It may safely be assumed that this is not what is observed. It would have 
shown up in the preliminary trials on the singles counts (see ref.~\cite{Wenger:2004}), 
which would have followed something suggestive of the basic predicted step 
function as the local oscillator phase shift was varied. What is observed in 
practice is more likely to be similar to the results obtained by Breitenbach 
\textit{et al.} \cite{Breitenbach:1995}. Their Fig.~6a, reproduced here as 
Fig.~\ref{fig4}, shows a distribution of photocurrents that is clustered around zero, 
for $\theta$ taking integer multiples of $\pi$, but is scattered fairly equally among 
positive and negative values in between. 

When digitised, the distribution would reduce to two straight horizontal 
lines, showing that for each choice of $\theta $ there is an equal chance of 
a `$+$' or a `$-$' result. As in any other Bell test setup, though, the absence 
of variations in the singles counts does not necessarily mean there is no variation in 
the coincidence rates.  As explained in the next section, however, the 
 coincidence curves are not the zig-zag ones of standard classical 
theory. These would be expected if we had \textit{full}
``rotational invariance''\footnote{``Rotational invariance'' means the hidden variable 
takes all possible values with equal probability.  In the current context, if the 
experiment does indeed produce high visibility coincidence curves, the hidden variable
responsible will be the common phase difference between test beams and local 
oscillators before application of the phase shifts (``detector settings'') 
$\theta_A$ and $\theta_B$.  It is argued that in the proposed experiment there will 
be at best \textit{approximate} rotational invariance, if there is appreciable 
variation in the (again common) frequency.}.  If the ideas presented here are 
correct, we have instead, in the language of an article by the 
present author \cite{Thompson:1999}, only \textit{binary} rotational invariance.  
Breitenbach's scatter of photocurrent differences is 
seen as evidence that the relative phase can (under perfect conditions) take 
just one of two values, 0 or $\pi$.  The scatter is formed from a superposition 
of two sets of points, corresponding to two sine curves that are out of phase, 
together with a considerable amount of noise.

This interpretation accords well with more comprehensive results of the experiment as
reported elsewhere \cite{Breitenbach:1997}.  When part of the initial laser beam is used
as ``seed'' to the OPA, judicious adjustments of the phase can produce ``bright squeezed
light'' and a scatter with alternately positive and negative points.  The presence of the 
seed causes selection of one particular phase set.

The two ``phase sets'' arise from the way in which the pulses are produced, 
which involves, after the frequency doubling, the \textit{degenerate} case of 
parametric down-conversion, the latter producing pulses that are 
(in contrast to the general case of conjugate frequencies) of 
\textit{exactly equal frequency}. Consider an initial pump laser of frequency 
$\omega$.  In the proposed experiment, this will be doubled in the crystal SHG 
to 2$\omega$ then halved in the OPA back to $\omega$.  At the frequency doubling 
stage, one laser input wave peak gives rise to two output ones. Assuming that there 
are causal mechanisms involved, it seems inevitable that every other wave peak 
of the output will be exactly in phase with the input. When we now use this 
input to produce a down-conversion, the outputs will be in phase either with 
the even or with the odd peaks of the input, which will make them either in 
phase or exactly out of phase with the original laser.  [The matter can alternatively 
be approached mathematically, as per ref.~\cite{Walls:1994}, where it is treated as
resonance in a nonlinear situation in which there are two solutions.]

We thus have two classes of output, differing in phase by $\pi$. If we 
define the random variable $\alpha$ to be 0 
for one class, $\pi$ for the other, this will clearly be an important 
``hidden variable'' of the system.

The existence of two classes of output of exactly equal frequency and exactly 
opposite phase may well be a feature common to a number of different experiments
employing degenerate parametric down-conversion sources. 
One example is discussed in ref~\cite{Thompson:1999},
namely the Bell test experiment conducted by Weihs \etal~\cite{Weihs:1998},
but there are many more and further examples, not all concerned 
with Bell tests, will be given in later papers.  Accepted theory is handicapped by
various pre-conceptions.  In some cases, for example ref.~\cite{Tittel:1998}, the 
assumption is made that even in the degenerate case the output pair have conjugate, not identical,
frequencies.  (If this is the case in the proposed experiment, though, it will severely 
reduce the visibility of any coincidence curve observed when the experimental beam is 
mixed back with the source laser in the homodyne detector.)
In other cases the use of the standard model in terms of orthogonal 
quadratures leads to neglect of more appropriate models.

As regards the possibility of any difference in frequency between the test beam
and the master laser, the preliminary experiments for the Garc\'{\i}a-Patr\'{o}n proposal 
\cite{Wenger:2004}, using just one output beam may already be sufficient to show that the interference 
is stronger than would then be the case.  It is known that the source laser has 
quite a broad band width, i.e.~that $\omega _{0}$ is not constant. Though it is likely that 
it is only part of the pump spectrum that induces a 
down-conversion, so that the band width of the test beam may be considerably 
narrower than that of the pump, it too is non zero. It follows that 
agreement of frequency between this and the test beam must be because we are 
always dealing, in the degenerate case, with \textit{exact} frequency doubling and 
halving.

\section{A classical model of the proposed Bell test}
In the proposed Bell test of Garc\'{\i}a-Patr\'{o}n  \textit{et al.~}, positive voltage 
differences will be treated as +1, negative as --1. Applying this version of 
homodyne detection 
to both branches of the experiment, the CHSH test ($ -2 \leq S \leq 2$) will 
then be applied to the coincidence counts. Under quantum theory it is 
expected that, so long as ``non-classical'' beams are employed, the test 
will be violated. However, since there are no obvious loopholes in the 
actual Bell test (see later), there is no apparent reason in our model
why local realism should not win: the test should 
\textit{not} be violated. In the classical view, this prediction is unrelated to any 
supposed non-classical nature of the light.

\subsection{The basic local realist model}
If we take the simplest possible case, in which to all intents and purposes 
all the frequencies involved are the same, the hidden variable in the local 
realist model is clearly going to be the phase difference ($\alpha = 0$ or $\pi$) 
between the test signal and the local oscillator.  If high visibility 
coincidence curves are seen, it must be because the values of $\alpha$ are identical for
the A and B beams.  Assuming no noise, the basic model is easily written down.

From equation (\ref{eq5}), the probability of a $-1$ outcome on side A is

\begin{equation}
\label{eq7}
p_{-} (\theta _{A}, \alpha )= \left\{
\begin{array}{ll}
    1 & \mbox{for $-\pi  <  \theta _{A} - \alpha  < 0$} \\
    0 & \mbox{for $ 0 < \theta _{A} - \alpha < \pi $},
\end{array}
\right.
\end{equation}

\noindent
where $\theta _{A}$ is the phase shift applied to the local oscillator A, 
$\alpha$ is the hidden variable and all angles are reduced modulo $2\pi$. 
Similarly, the probability of a +1 outcome is

\begin{equation}
\label{eq8}
p_{+} (\theta _{A}, \alpha )= \left\{
\begin{array}{ll}
    0 & \mbox{for $-\pi  <  \theta _{A} - \alpha  < 0$} \\
    1 & \mbox{for $ 0 < \theta _{A} - \alpha < \pi $},
\end{array}
\right.
\end{equation}

Assuming equal probability $\frac{1}{2}$ for each of the two possible values of 
$\alpha$, the standard ``local realist'' assumption that independent 
probabilities can be multiplied to give coincidence ones leads to a 
predicted coincidence rate of
\begin{eqnarray}
\label{eq9}
P_{++} (\theta _A ,\theta _B)& = & \frac{1}{2} p_+ (\theta _A ,0)p_+ (\theta _B 
,0)
\nonumber \\
                        & + & \frac{1}{2} p_+ (\theta _A ,\pi )p_+ (\theta _B ,\pi ),
\end{eqnarray}
with similar expressions for $P_{+-}$, $P_{-+ }$ and $P_{- -}$. 

\noindent The result for $\theta _{A}=\pi/2$, for example, is
\begin{equation}
\label{eq10}
P_{++} (\pi /2, \theta _{B}) = \left\{
\begin{array}{ll}
     0 &  \mbox{for $-\pi  <  \theta _{B} <  0$}\\
   1/2 & \mbox{for $0 < \theta _{B} < \pi$}.
\end{array}
\right.
\end{equation}
For $\theta _{A}$ = --$\pi $/2 it is
\begin{equation}
\label{eq11}
P_{++} (-\pi /2, \theta _{B}) = \left\{
\begin{array}{ll}
     1/2 &  \mbox{for $-\pi  <  \theta _{B} <  0$}\\
       0 & \mbox{for $0 < \theta _{B} < \pi$}.
\end{array}
\right.
\end{equation}

\begin{figure}[htbp]
%\centerline{\includegraphics[width=5.99in,height=9.26in]{ThompsonFig5.eps}}
\centerline{\includegraphics[width=1.8in,height=2.7in]{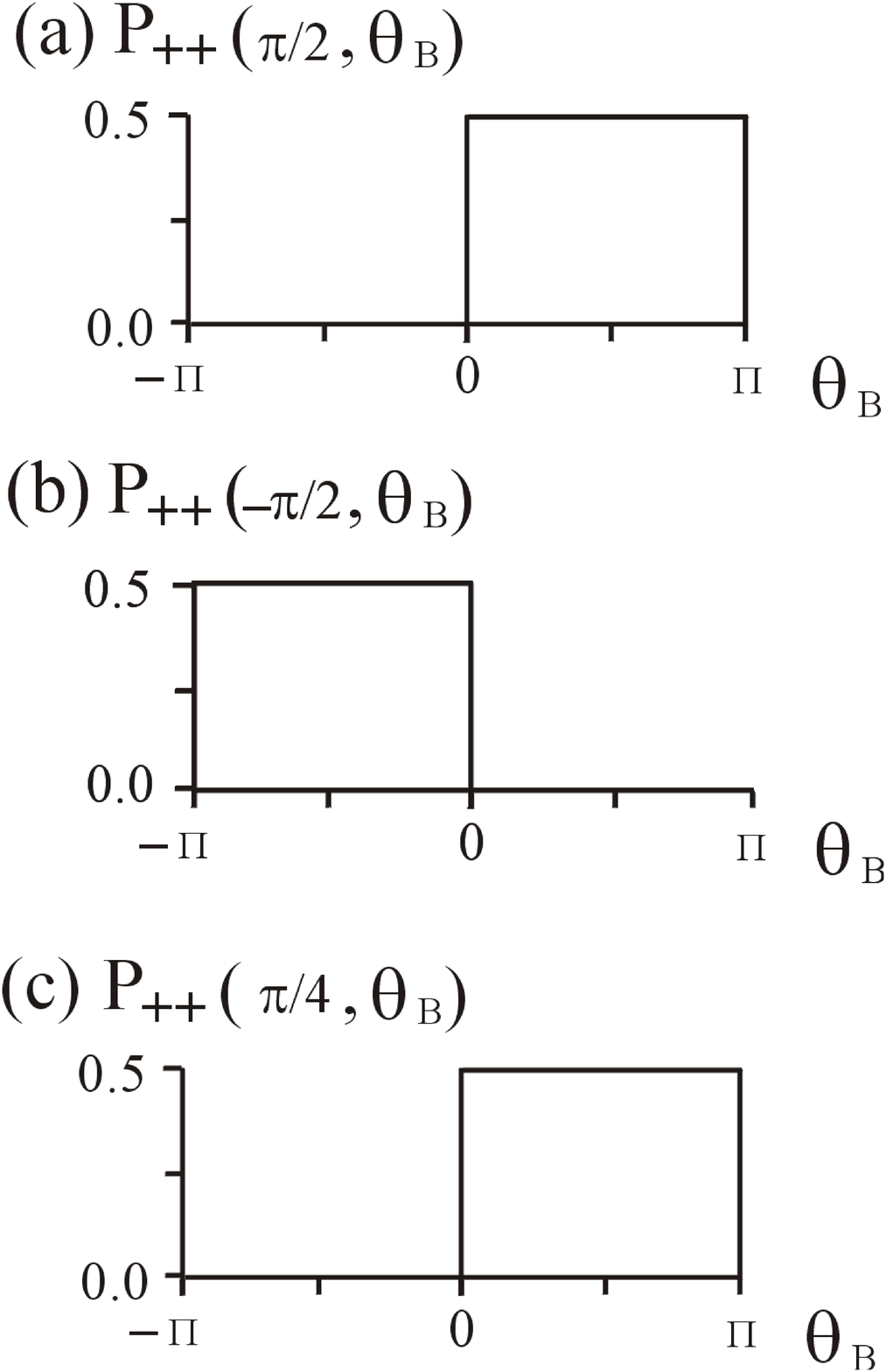}}

\caption{Predicted coincidence curves for the ideal experiment.  
(a) and (b) illustrate the settings most likely to be chosen in practice, 
giving the strongest correlations. $\theta _{A}$ is fixed at $\pi/2$ or 
$-\pi/2$ while $\theta _{B }$ varies. In theory, any value of $\theta 
_{A}$ between 0 and $\pi$ would give the same curve as (a), any between 
$-\pi$ and 0 the same as (b). An example is shown in (c), where $\theta 
_{A }$ is $\pi/4$ but the curve is identical to (a). We do not have 
rotational invariance: the curve is not a function of $\theta _{B }-\theta _{A}$.
}
\label{fig5}
\end{figure}

Note that, as illustrated in Fig.~\ref{fig5}, the coincidence probabilities \textit{cannot}, in this 
basic model, be expressed as functions of the difference in detector 
settings, $\theta _{B}- \theta _{A}$. This failure, marking a significant deviation 
from the quantum mechanical prediction, is an inevitable consequence of the fact that 
we have (as mentioned earlier) only binary, not full, rotational invariance.

\subsection{Fine-tuning the model}
Many practical considerations mean that the final local realist prediction 
will probably not look much like the above step function. It may not even be 
quite periodic. The main logical difference is that, despite all that has 
been said so far, the actual variable that is set for the local oscillators 
is not directly the phase shift but the path length, and, since the 
frequency is likely to vary slightly from one signal to the next (though 
always keeping the same as that of the pump laser), the actual phase 
difference between test and local oscillator beams will depend on the path 
length difference \textit{and} on the frequency. In a complete model, therefore, the 
important parameters will be path length and frequency, with phase derived 
from these.

If frequency variations are sufficiently large, the situation may approach 
one of rotational invariance (RI), but it seems on the face of it unlikely 
that this can be achieved without loss of correlation. If we do have RI, 
perhaps produced artificially by introducing random phase variations into the
OPA pump beam, the model becomes the standard realist one in which the predicted quantum 
correlation varies linearly with difference in phase settings, but it is 
more likely that what will be found is curves that are \textit{not} independent of the 
choice of individual phase setting. They will be basically the predicted 
step functions but converted to curves as the result of the addition of 
noise.

\begin{figure}[htbp]
%\centerline{\includegraphics[width=5.39in,height=2.80in]{ThompsonFig6.eps}}
\centerline{\includegraphics[width=2.0in,height=1.0in]{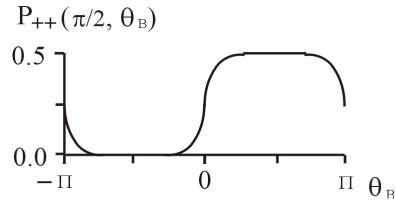}}
\caption{Likely appearance of coincidence curves in a real experiment with 
moderate noise.}
\label{fig6}
\end{figure}

It is essential to know the actual experimental conditions. Several relevant 
factors can be discovered by careful analysis of the variations in the raw 
voltages in each homodyne detection system. If noise is low, the presence of 
the two phase sets, and whether or not they are equally represented, should 
become apparent. 

All this complexity, though, has no bearing on the basic fact of the existence 
of a hidden variable model and the consequent prediction that the CHSH Bell test 
will not be violated.

\subsection{The role of the ``event-ready'' detectors}
In the quantum-mechanical theory, the expectation of violation of the Bell 
test all hinges on the production of ``non-classical'' light. The light 
directly output from the crystal OPA is assumed to be Gaussian, i.e.~it 
takes the form of pulses of light that have a Gaussian intensity profile and 
also, as a result of Fourier theory, a Gaussian spectrum. When this is 
passed through an unbalanced beamsplitter (BS$_{A}$ or BS$_{B}$) and a 
``single photon'' detected by an ``event-ready'' detector, the theory says 
that the subtraction of one photon leaves the remaining beam 
``non-Gaussian''. Although there is mention here of single photons, the 
theory is concerned with the ensemble properties of the complete beams, not 
with the individual properties of its constituent pulses.

In the local realist (classical) model considered here, the shapes of the spectra 
are not relevant except insofar as a narrow band width is desirable for the 
demonstration of dramatic correlations. The ``event-ready detectors'' (PD$_{A}$ 
and PD$_{B}$ in Fig.~\ref{fig1}) play, instead, the important role of selecting 
for analysis only the strongest down-converted output pairs, it being assumed that 
the properties of the transmitted and reflected light at the unbalanced beamsplitters 
are identical apart from their amplitudes.  It is likely that those detected signals 
that are coincident with each other will be genuine ``degenerate'' ones, 
i.e.~of exactly equal frequency, quasi-resonant with the pump laser. The unbalanced 
beamsplitters and the detectors PD$_{A}$ and PD$_{B}$ need to be set so that 
the intensity of the detected part is sufficient to be above the minimum for 
detection but low enough to ensure that all but the strongest pulses are 
ignored.

In neither theory are the event-ready detectors really needed in their 
``Bell test'' role of ensuring a fair test (see below), since the homodyne 
detectors are almost 100{\%} efficient.

\section{Validity of the proposed Bell test}
Coincidences between the digitised voltage differences will be used in the 
CHSH Bell test \cite{Clauser:1969, Thompson:1996}, but avoiding the ``post-selection'' 
that has, since 
Aspect's second experiment \cite{Aspect:1982}, become customary.
The Garc\'{\i}a-Patr\'{o}n \textit{et al.}~proposal is to use event-ready detectors, 
as recommended by Bell himself for use in real experiments \cite{Clauser:1978}.
None of the usual loopholes \cite{Thompson:2003} are expected to be applicable:

\begin{enumerate}
\item With the use of the event-ready detectors, non-detections are of little concern. 
The detectors (PD$_{A}$ and PD$_{B}$ in Fig.~\ref{fig1}) act to define the sample to be analysed, 
and the fact that they do so quite independently of whether or not any member of the sample 
is then also detected in coincidence at the homodyne detectors ensures that no bias is 
introduced here. The estimate of ``quantum correlation''\footnote{In the derivation of Bell inequalities such as the CHSH 
inequality, the statistic required is simply the expectation value of the product of the 
outcomes.  This is commonly referred to in this context as the ``quantum correlation''. 
Only when there are no null outcomes, possible outcomes being restricted to just $+1$ 
or $-1$, does it coincide with ordinary statistical correlation.} 
to be used in calculating the 
CHSH test statistic is $E = (N_{++AB} + N_{--AB} - N_{+-AB} - N_{-+AB}) / N_{AB}$, where 
the $N$'s are coincidence 
counts and the subscripts are self-explanatory. This contrasts with the usual method, in 
which the denominator used is not $N_{AB}$ but the sum of observed coincidences, 
$N_{++AB} + N_{--AB} + N_{+-AB} + N_{-+AB}$. The use of the latter can readily be shown 
to introduce bias unless it can be assumed that the sample of detected pairs is a fair one. 
That such an assumption fails in some plausible local realist models has been known 
since 1970 or earlier \cite{Thompson:1996, Pearle:1970}. 
\item There is no possibility of synchronisation problems \cite{Fine:1982}, since a pulsed source is used.
\item No ``accidentals'' will be subtracted \cite{Thompson:2003}. 
\item The ``locality'' loophole can be closed by using long paths and a random system for 
choosing the ``detector settings'' (local oscillator phase shifts) during the propagation of 
the signals.
\end{enumerate}

The system is almost certainly not going to be ``rotationally invariant'' 
(not all phase differences will be equally likely), but this will not 
invalidate the Bell test. It may, however, be important in another way. It 
is likely that high visibilities will be observed in the coincidence curves 
(i.e.~high values of (max -- min)/(max + min) in plots of coincidence rate 
against difference in phase shift), leading to the impression that the Bell 
test ought to be violated. These visibilities, though, will depend on 
the absolute values of the individual settings. High ones will be balanced by low, 
with the net effect that violation does not in fact happen.

\section{Validity and significance of negative estimates for Wigner densities}
\begin{figure}[htbp]
%\centerline{\includegraphics[width=5.33in,height=4.00in]{ThompsonFig7.eps}}
\centerline{\includegraphics[width=2.7in,height=2.00in]{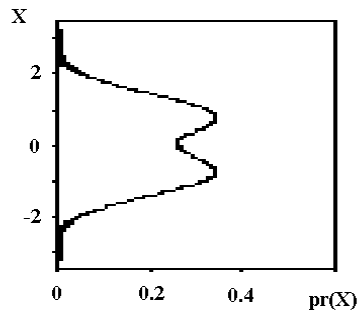}}
\caption{Observed distribution of voltage differences, X, using homodyne 
detection in an experiment similar to the proposed Bell test and averaging 
over a large number of applied phase shifts. (\textit{Based on Fig.~4a of 
A.~I.~Lvovsky et al., Phys.~Rev.~Lett.~} \textbf{\textit{87}}\textit{, 050402 (2001).})}
\label{fig7}
\end{figure}

In the quantum mechanical theory discussed in the loophole-free Bell test 
proposals and in other recent papers \cite{Lvovsky:2001,Wenger:2004}, 
part of the evidence that is put forward as indicating that negative Wigner 
densities are likely to be obtained consists in the observation that, when 
$\theta$ is varied randomly, the distribution of observed voltage differences 
shows a double peak (see Fig.~\ref{fig7}). There is a tendency to observe 
roughly equal numbers of + and -- results but relatively few near zero. The 
fact that the relationship depends on the sine of $\theta$ is, however, 
sufficient to explain why this should be so. 

To illustrate, let us consider the following. The sine of an angle is 
between 0 and 0.5 whenever the angle is between 0 and $\pi/6$. It is 
between 0.5 and 1.0 when the angle is between $\pi/6$ and $\pi/2$. Since 
the second range of angles is twice the first yet the range of the values of 
the sine is the same, it follows that if all angles in the range 0 to $\pi/2$ 
are selected equally often there will be twice as many sine values seen 
above 0.5 as below.  The same holds when random angles between $\pi/2$ and $\pi$
are chosen, whilst for values between $-\pi$ and 0 we find a symmetrical result for
negative values.

When allowance is made for the addition of noise, the production of a 
distribution such as that of Fig.~\ref{fig7} for the average when angles are sampled 
uniformly comes as no surprise. Clearly, as the experimenters themselves 
recognise, the dip is not in itself sufficient to prove the non-classical 
state of the light. For this, direct measurement of the Wigner density is 
required, but there is a problem here. No actual direct measurement is 
possible, so it has to be estimated, and the method proposed is the Radon 
transformation \cite{Leonhardt:1997}. It is claimed that in other experiments 
Wigner densities calculated either by this procedure or by a ``more precise 
maximum-liklihood reconstruction technique'' \cite{Babichev:2004} 
have shown negative regions, but perhaps the methods should be checked for 
validity?

In any event, as already explained, the natural hidden variable relevant to 
the proposed experiment is the phase of the individual pulse, not any 
statistical property of the whole ensemble.  Indeed, the use of Wigner density as
a substitute for hidden variables was never originally intended and has no
theoretical basis.  In Bell's much-quoted paper on the subject (Ch.~21 of 
ref.~\cite{Bell:1987}), the ``hidden variables'' remain, as ever, parameters such 
as position and momentum that are specific to individual particles.  The role 
of the negative Wigner density is merely to provide, in certain rather special 
circumstances, and alternative test for nonlocality in that, if negative values 
are found, then real local hidden variables cannot exist.

\section{Suggestions for extending the experiment}
The basic set-up would seem to present an ideal opportunity for investigation of some key 
aspects of the quantum and classical models, as well as the operation of the 
Bell test ``detection loophole''. 

\begin{enumerate}
\item \textbf{The operation of the detection loophole} could be illustrated if, 
instead of using the digitised difference voltages of the homodyne 
detectors, the two separate voltages are passed through discriminators. The 
latter operate by applying a threshold voltage that can be set by the 
experimenter and counting those pulses that exceed it. These can be used in 
a conventional CHSH Bell test, i.e.~using total observed coincidence count as 
denominator in the estimated quantum correlations $E$. 
The model that has been known since Pearle's time (1970) predicts that, as 
the threshold voltage used in the discriminators is increased and hence the 
number of registered events decreased (interpreted in quantum theory as the 
detection efficiency being decreased), the CHSH test statistic $S$, if calculated 
using estimates $E = (N_{++} + N_{--} - N_{+-} - N_{-+}) / 
(N_{++} + N_{--} + N_{+-} + N_{-+})$, will increase. 
If noise levels are low, it may well exceed the Bell limit of 2.

Such an experiment has, in a sense, already been performed by Babichev \etal
\cite{Babichev:2004} and yielded the expected results.  If it is accepted that their
source (two outputs from a beamsplitter) would have been in an entangled state, then
their Fig.~4b clearly demonstrates that high detector thresholds (i.e.~low detector 
efficiencies) can lead to violations of the standard form of the CHSH test.

\item \textbf{The existence of the two phase sets} is, in point of fact, well known when
OPA's are operated above threshold \cite{Walls:1994}.  The resonance into one or 
other of the sets then becomes stable.  The fact that that two sets are also 
responsible for many of the observations below threshold could be further investigated if 
either the raw voltages or the undigitised difference voltages are analysed. 
So long as the noise level is low, the existence of the two superposed 
curves, one for $\alpha = 0$ and the other for $\alpha =\pi$, should 
be apparent.  It would be interesting to investigate how the pattern changed
as optical path lengths were varied.  Breitenbach's pattern might be hard to 
reproduce using long path lengths, where exact equality is needed unless the
light is monochromatic.

\item \textbf{Comparison of overall performance:} If the primary goal of the 
experimenter is clearly set out to be the comparison of the performance of the 
two rival models, rather than merely the conduct of a Bell test, further ideas 
for modifying the set-up will doubtless emerge when the first experiments have been 
done.  Many of the predictions of the quantum-mechanical model have already appeared 
in print \cite{Garc:2004,Nha:2004}.  The first stage in comparing models should 
probably be, therefore, to conduct supplementary experiments so as to
establish the relevant parameters of the full classical model and hence make
equivalent empirical predictions.  It is possible, though, that qualitative predictions
alone will be sufficient to demonstrate superiority one way or the other.
\end{enumerate}

\section{Conclusion}
The proposed experiments would, \textit{if the ``non-classicality'' of the 
light could be demonstrated satisfactorily}, provide a definite answer one way or the 
other regarding the reality of quantum entanglement. They could usefully be 
extended to include empirical investigations into the operation of the Bell 
test detection loophole. Perhaps more importantly, though, they present 
valuable opportunities to compare the performance of the two theories in 
both their total predictive power and their comprehensibility. Are 
parameters such as ``Wigner density'' and ``degree of squeezing'' really the 
relevant ones, or would we gain more insight into the situation by talking 
only of frequencies, phases and intensities? Parameters such as the 
detection efficiency and the transmittance of the beamsplitters will 
undoubtedly affect the results, but do they do this in the way the quantum 
theoretical model suggests? It will take considerably more than just the 
minimum runs needed for the Bell test if we are to find the answers.

The detailed predictions of the classical model cannot be given until 
the full facts of the experimental set-up and the performance of the various 
parts are known, but it gives, in any event, a simple explanation of the 
double-peaked nature of the distribution of voltage differences. The peaks 
arise naturally from the way in which homodyne detection works, and the 
quantum theoretical idea that they are one of the indications of a non-classical beam 
or of negative Wigner density would not appear to be justifiable. The idea 
that a classical beam can become non-classical by the act of ``subtracting 
a photon'' is, equally, of doubtful validity.  The experimental role of the 
subtraction and detection of part of each beam is to aid the selection for 
coincidence analysis of those pulses that are likely to be most strongly correlated.

\ack{Acknowledgements}
I am grateful to Ph.~Grangier for drawing my attention to his team's 
proposed experiment \cite{Garc:2004}, and for helpful discussions.  I should also
like to thank G.~Breitenbach and H.~Carmichael for pointing me in
the direction of related work.

\appendix
\section*{Appendix. Alternative classical derivation of the homodyne detection 
formula}
\setcounter{section}{1}
A derivation is given here that does not involve complex numbers and hence 
confirms that the equations in the text are ordinary wave equations, not 
quantum-mechanical ``wave functions''. 

The relationship between intensity 
difference and the local oscillator phase can be checked as follows:

\bigskip
\noindent Assume the two input beams are
\medskip

\par Experimental beam: $E \cos \phi$ , where, as before, $\phi =\omega t$
\nopagebreak 
\par Local oscillator: $E_{L} \cos (\phi +\theta )$

\medskip
\noindent Then the output beams, assuming a 50-50 beamsplitter and no losses, can be 
written
\medskip

Reflected beam: 
\begin{eqnarray}
\label{eqA1}
E_r & = & \frac{1}{\sqrt 2} (E\cos (\phi + \pi /2) + E_L \cos (\phi + \theta))
\nonumber \\
    & = & \frac{1}{\sqrt 2} (-E\sin \phi + E_L \cos (\phi + \theta))
\end{eqnarray}

Transmitted beam: 
\begin{eqnarray}
\label{eqA2}
E_t  & = & \frac{1}{\sqrt 2} (E\cos \phi + E_L \cos (\phi + \theta + \pi /2))
\nonumber \\
     & = & \frac{1}{\sqrt 2} (E\cos \phi - E_L \sin (\phi + \theta)).
\end{eqnarray}

\noindent Let us define a (constant) angle $\psi$ such that $\tan \psi = E / E_{L}$,
 making $E \cos \psi = E_{L } \sin \psi$.

\medskip
\noindent
Consider the case when $\theta = 0$.  We have

\begin{eqnarray}
\label{eqA3}
E_r & = & \frac{1}{\sqrt 2} (E/\sin \psi )(-\sin \psi \sin \phi +\cos \psi \cos \phi )
\nonumber \\
    & = & \frac{1}{\sqrt 2} (E/\sin \psi )\cos (\psi +\phi ),
\end{eqnarray}
\noindent
so that amplitude is proportional to $\frac{1}{\sqrt 2} E / \sin \psi$ and 
intensity to $\frac{1}{2} E^{2} / \sin^{2} \psi$.

The intensity of the transmitted beam can be found similarly, and turns out to be 
also proportional to $\frac{1}{2} E^{2} / \sin^{2} \psi$, so that the voltage 
difference from the homodyne detector is therefore to be zero.   Likewise, a zero 
difference is found for $\theta =\pi$, but other values of $\theta$ produce more 
interesting results.
\medskip 

\noindent For example, for $\theta =\pi /2$ we find
\begin{eqnarray}
\label{eqA4}
E_r & = & \frac{1}{\sqrt 2} (E/\sin \psi )(-\sin \psi \sin \phi - \cos \psi \sin \phi )
\nonumber \\
    & = & \frac{1}{\sqrt 2} (E/\sin \psi )\sin \phi (-\sin \psi -\cos \psi ),
\end{eqnarray}
\begin{eqnarray}
\label{eqA5}
E_t & = & \frac{1}{\sqrt 2} (E/\sin \psi )(\sin \psi \cos \phi -\cos \psi \cos \phi )
\nonumber \\
    & = & \frac{1}{\sqrt 2} (E/\sin \psi )\cos \phi (\sin \psi -\cos \psi ).
\end{eqnarray}

\noindent The difference in intensities is therefore proportional to 
\begin{eqnarray}
\label{eqA6}
\mbox{Difference} & = & \frac{1}{2} (E^2/\sin ^2\psi )[(-\sin \psi - \cos \psi )^2
\nonumber \\
       & - & (\sin \psi - \cos \psi )^2]
\nonumber \\
       & = & \frac{1}{2} (E^2/\sin ^2\psi )4\sin \psi \cos \psi )
\nonumber \\
       & = & \frac{1}{2} E^2\cos \psi /\sin \psi 
\nonumber \\
       & = & 2E^2E_L /E
\nonumber \\
       & = & 2EE_L .
\end{eqnarray}

This is consistent with the result obtained by the method in the main text, 
so it seems safe to accept that as being correct.

\end{document}